\documentclass[conference]{IEEEtran}
\IEEEoverridecommandlockouts
\usepackage{cite}
\usepackage{amsmath,amssymb,amsfonts}
\usepackage{algorithmic}
\usepackage{graphicx}
\usepackage{textcomp}
\usepackage{xcolor}
\usepackage{verbatim}
\usepackage{multirow}
\usepackage{footmisc}
\usepackage{fancyvrb}
\usepackage{framed}
\usepackage{subcaption}
\usepackage{ulem}
\usepackage{todonotes}

\newcommand{\redtext}[1]{\textcolor{red}{#1}}

\def\BibTeX{{\rm B\kern-.05em{\sc i\kern-.025em b}\kern-.08em
    T\kern-.1667em\lower.7ex\hbox{E}\kern-.125emX}}
    
\begin{document}

\title{Strengthening SDN Security: \\ Protocol Dialecting and Downgrade Attacks}


\author{Michael Sjoholmsierchio~~~Britta Hale~~~Daniel Lukaszewski~~~Geoffrey G.\ Xie \\[0.05in] Naval Postgraduate School
\\
[0.05in] October 20, 2020
}

\maketitle
\thispagestyle{plain}
\pagestyle{plain}

\begin{abstract} 
Software-defined networking (SDN) has become a fundamental technology for data centers and 5G networks. In an SDN network, routing and traffic management decisions are made by a centralized controller and communicated to switches via a control channel. Transport Layer Security (TLS) has been proposed as its single security layer; however, use of TLS is optional and connections are still vulnerable to downgrade attacks.  In this paper, we propose the strengthening of security assurance using a protocol dialecting approach to provide additional and customizable security. We consider and evaluate two dialecting approaches for OpenFlow protocol operation, adding per-message authentication to the SDN control channel that is independent of TLS and provides robustness against downgrade attacks in the optional case of TLS implementation. Furthermore, we measure the performance impact of using these dialecting primitives in a Mininet experiment. The results show a modest increase of communication latency of less than 22\%.
\end{abstract}

\vspace{0.1in}
\begin{IEEEkeywords}
Network security, Software-Defined Networks, Protocol Dialect, Transport Layer Security
\end{IEEEkeywords}

\normalem

\section{Introduction}
\label{sec:intro}

In contrast to \mbox{traditional} IP data networks, software-defined networks (SDNs) centralize  network control functions such as routing to a programmable decision element -- the controller. This enables the separation of the control functions (which we collectively referred to as the control channel) from the data channel functions that forward actual user packets.  The controller provides a single platform for programming and orchestrating network control functions to  respond to user traffic demands based on a network-wide view of current network state~\cite{greenberg05}.  Devices operating in the SDN data channel (called switches) simply accept and enforce decisions (in the form of new flow rules) provided from the controller via the control channel.  The separation of control channel and data channel enables streamlined network administration, fine-grain flow-level security enforcement~\cite{ethane07}, resource adaptability~\cite{hedera10}, and scalability~\cite{kim13}. SDNs have become a fundamental technology for data centers and 5G networks~\cite{cui16}, where resource efficiency and application performance requirements are more stringent than traditional data networks. 

\begin{figure}[!htbp]
    \centering
    \includegraphics[width=80mm]{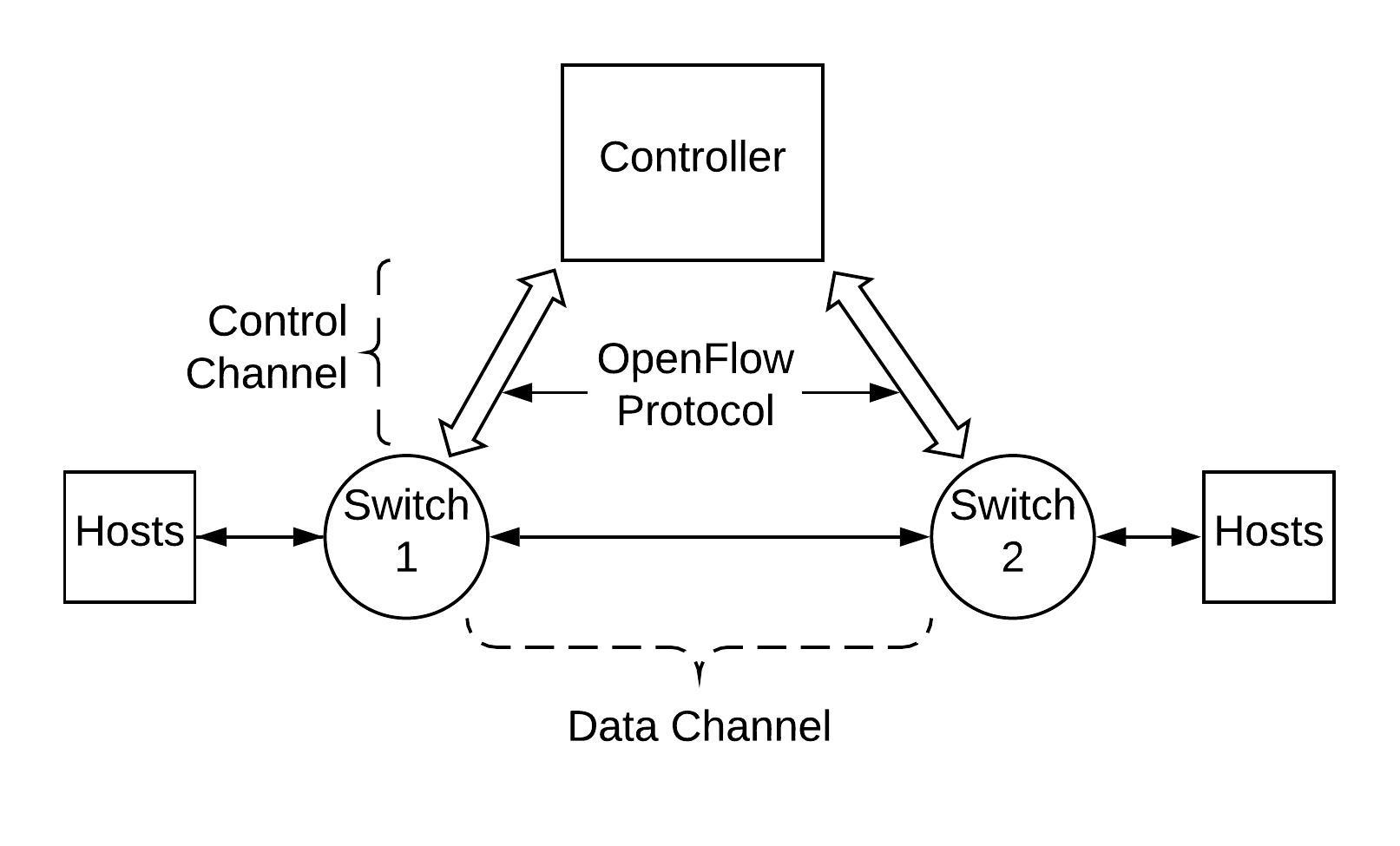}
    \caption{Overview of SDN Network. }
    \label{fig:SDN-network}
\end{figure}

SDNs come with two interfaces for a controller to interact with other network components~\cite{of1.5}. The \textit{northbound} interface is a controller software platform-specific application programming interface (API) for a unified communication model with deployed network control applications (traffic engineering, access control, etc). For example, the ONOS controller software\cite{onos} is able to support and orchestrate a wide range of network control functions while using a distributed software architecture to enhance fault tolerance.  The \textit{southbound} interface is used to support exchange of control messages between a controller and switches in the same network and is currently carried out by the standard OpenFlow protocol~\cite{of1.5}. OpenFlow runs over TCP and is the only means for a controller to communicate with SDN capable switches regardless of the vendor, making OpenFlow the exclusive standardized protocol used by the SDN control channel.

The control channel carries all decisions and commands for packet forwarding, switch configuration settings, and security operations from the controller to the switches. Correspondingly, network interface state updates and packet forwarding decision requests are relayed from switches to the controller. Thus, we refer to the controller and switches as \emph{SDN devices}, while hosts are considered as external to the SDN network and include endpoint devices such as clients/servers not participating in OpenFlow communications. The basic SDN configuration is shown  in Fig. \ref{fig:SDN-network}. Given the increased importance of SDNs and the critical role of the control channel to these networks, security of the SDN control channel is critical.

Currently, Transport Layer Security (TLS) protocol is the only recommended security solution for protecting the SDN control channel OpenFlow traffic, and it is only considered on an optional basis~\cite{of1.5}. TLS, if enabled, provides a layer of authentication and confidentiality. There are two main challenges of implementing TLS with OpenFlow. First, because TLS is optional and not a requirement, not all commercial SDN switch vendors and controller platforms provide native support of TLS \cite{of1.5}. Secondly, TLS is vulnerable to downgrade attacks~\cite{rfc7457,downgrade,ncc19}. Downgrade attacks are able to trick the communicating partners in the initial TLS handshake phase to adopt a potentially vulnerable protocol version or weaker ciphersuites (i.e. version downgrade and ciphersuite downgrade attacks). Such an attack would leave command data, such as flow rules, vulnerable to modification before being implemented by a switch.

The IETF has worked to develop solutions to address downgrade attack issues, e.g.,~\cite{rfc8446}. However, use of new TLS versions can require time for wide deployment, particularly in SDNs, and are still only provided on an optional basis.  In this paper, following the defense-in-depth principle, we investigate \textit{a complementary solution using} a novel protocol dialecting approach.  Protocol dialecting is an emerging network operation technique that aims to specialize the communication syntax and semantics of a standard networking protocol for the purpose of enhancing network security~\cite{xu19,mao19}. The addition of a dialect also requires no action or modification of the protocol standard that it is added to. Modifications to existing protocols to achieve new goals form the roots of protocol dialecting.  This practice of experimenting with existing or optional fields of a protocol to signal a desired behavior is not new.  The standard TCP protocol has incorporated multiple extensions through the use of the TCP Option field, to include window scaling, fast open, and timestamps. For instance, the creation of the Multipath TCP protocol takes advantage of the root TCP protocol by adding in a Multipath Capable option~\cite{paasch}. The TCP 3-way handshake can now provide a signal that the user wishes to establish multiple TCP connections with the server, and fall back to normal TCP behavior if the option is not supported.

It might seem straightforward to add new security checks for specific purposes. The novelty of this work lies largely in the combination of the new features and security layers with existing security solutions (such as TLS) in a principled way. We demonstrate the addition of novel security features that may be used separately or in conjunction with the optional TLS, and demonstrate management of additional complexity to network operations.

Our contributions are as follows:
\begin{enumerate}
\itemsep 0.1in
  \item We formulate high level criteria and a methodology for systematically designing and evaluating a protocol dialect. 
  
  \item We design and evaluate two security \mbox{derivatives} (potential dialecting solutions) that can be used to dialect the OpenFlow protocol. We present the security rationale for the derivatives to achieve another layer of per-message authentication that is independent of TLS, and as such, enhance the defense of some TLS versions against certain attacks. We further present experimental results to show that deploying the derivatives will likely incur a modest increase of communication latency.   
  
  \item We conceptualize a new form of policy-based networking by extending the protocol dialecting functionality at a SDN device into a policy enforcement proxy (PEP). 
  
  \item We simulate a downgrade attack and demonstrate how the designed and implemented  dialect could be used to detect and prevent such attacks against a Mininet-emulated SDN environment.
\end{enumerate}

\vskip 0.05in

As a direct consequence of our third contribution,  an operator could leverage the proxies for optionally defining and enforcing security-oriented policy for specific protocols, e.g., by mandating the use of TLS or permitting only specific TLS versions for the SDN control channel. A PEP is used as the implementation method for our experiment; however, protocol dialects are not limited to use of proxies.

The paper is organized as follows. In Section~\ref{sec:preliminaries}, we ground the work with an overview of the protocol dialecting concept, a systematic design methodology, a  description of TLS downgrade attacks, and a review of related work. In Section~\ref{sec:method}, we present the design of two derivatives for dialecting the OpenFlow protocol that add a layer of message authentication which may optionally be used in conjunction with TLS. Section~\ref{sec:analysis} details the security rationale for our design and \mbox{Section~\ref{sec:evaluation}} describes results from our Mininet experimentation with the derivatives. Finally, we discuss limitations and potential extensions of this work in Section~\ref{sec:discussion} and offer concluding remarks in Section~\ref{sec:conclusion}.

\section{Preliminaries}
\label{sec:preliminaries}

\begin{figure*}[th]
{\centering
  \includegraphics[width=7.2in]
    {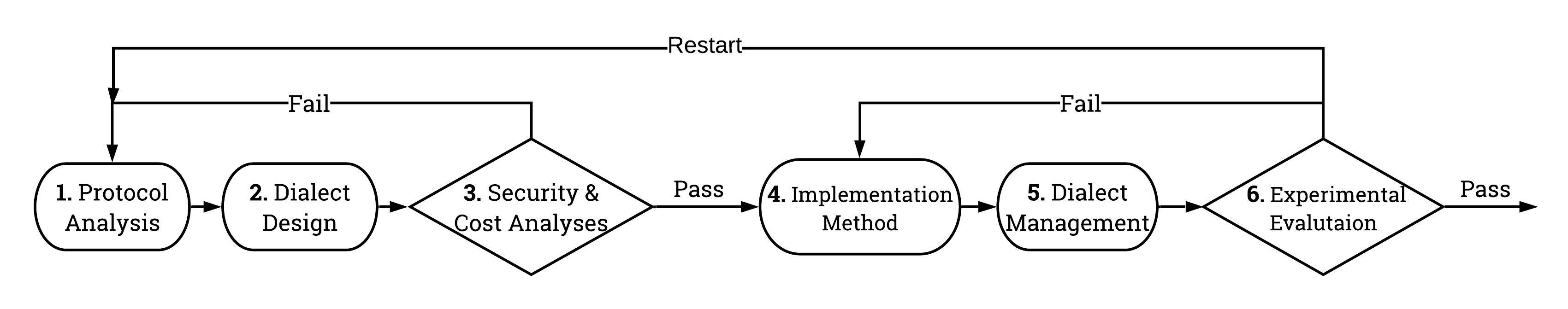}
  \caption{6-step Methodology for Designing and Evaluating a Protocol Dialect. 
  }
  \label{fig:method}
}
\end{figure*}
The concept of protocol dialecting, as well as its use as a network security solution, has been applied to a relatively small number of networking protocols. In this section, we provide a brief overview of the technical area and a formulation for systematically designing and evaluating a protocol dialect. We also detail the steps of a downgrade attack against TLS to substantiate the claim that TLS by itself has limitations in securing SDN control channels. Finally, we review prior work that is closely related to our proposal.

\subsection{Protocol Dialecting} 

Protocol dialecting aims to allow an operator to \emph{\mbox{specialize} operation of a networking protocol beyond the standard \mbox{configuration} options}, potentially on a per-network basis. In today's data network operation, it is common to use open-source implementations of standard protocols such as DNS, HTTPS, and OpenFlow. Therefore, such networks share the same software vulnerabilities and cannot escape from the dreadful reality of ``break one [network], break them all''. By adding network-specific protocol behaviors, the premise of protocol dialecting is to \emph{force potential attackers to spend significantly more resources} to decipher specific details of the protocol specialization and to meet customized security needs of an organization. The goal of the particular dialecting approach presented here is to cause an attacker to spend more time and resources compared to networks with only TLS, as well as to enable per-message authentication secrets.

Typically, a protocol dialect employs multiple underlying sub-solutions for customizing the operation of a protocol. Each of the sub-solutions is referred to as \emph{a dialecting derivative} in this paper. Derivatives can take many forms.
In reported use cases, some researchers have proposed to change a protocol's messaging semantics (e.g., re-purposing of fields and private options) and timing (e.g., artificial delays, duplicates, or omissions of messages)~\cite{xu19,mao19}. Direct modifications of protocol semantics and timing have the additional benefit of amplifying the presence of attack traffic, in the same way that a visitor who doesn't speak the local dialect would stand out when speaking to a roomful of locals. In one of our derivatives, we follow this precedence and require no message modification; in the second derivative we consider modifications to the format and semantics of OpenFlow messages. The first derivative adds security into existing data packets using a normally random packet field, while the second derivative provides security as a wrapper around the entire unmodified packet.

\subsection{Toward Systematic Development of Protocol Dialects} 
\label{sec:criteria}

Protocol dialecting must be friendly to network operators to ensure that its implementation does not greatly increase network management complexity. We observe that two factors are critical to controlling the complexity. First, most network operators do not have the programming or security background required for safely modifying protocol software, even if the source code is available.  Automated tools for creating protocol dialects from original binaries is highly desirable. Alternatively, plug-and-play proxy software with minimum configuration requirements can be deployed \sout{between devices} to alter protocol behaviors without the need to modify protocol binaries. The use of proxy software reduces the requirement of tailoring to the underlying protocol being modified and can be protocol agnostic. Second, protocol dialects should not adversely affect the effectiveness of current network monitoring tools. This means that modifications made to the OpenFlow packets or  wrappers do not modify the OpenFlow protocol and can be removed prior to analysis by a monitoring tool. The resulting traffic flows should not cause significant manual re-configurations of intrusion detection systems. 

From the above observations, we propose these two criteria for properly evaluating the effectiveness of a protocol dialect: 
\begin{enumerate}
 \item {\bf Security benefits}. The protocol dialect must provide net security benefits in that it enables significantly improved security protection without introducing new vulnerabilities that could weaken the security properties of the original protocol implementation.
 \item {\bf Deployment costs}. The expertise, time, and resources required of the network manager to deploy the protocol dialect must be at a similar level to that in a typical deployment of the original protocol implementation. Similarly, deployment of the protocol dialect must not significantly increase network communication overhead.
\end{enumerate}
Next, we present a methodology for systematically designing, testing, and deploying a protocol dialect to ensure meeting both criteria. The methodology consists of six distinct stages as illustrated in Fig.~\ref{fig:method}.

Central to our methodology are two iterative processes specifically designed to mandate that both aforementioned evaluation criteria are considered. The process in the left half of Fig.~\ref{fig:method} evaluates the security benefits and deployment cost primarily through formal modeling and analysis while refining assumptions about the targeted deployment scenarios. Ideally, the dialect provides a general solution with sufficient flexibility to work with networks of varying size, topology, and services provided. Assumptions must explicitly, and precisely, characterize resource and time constraints, threat model, and strengths and weaknesses of available security solution building blocks. These assumptions will likely go through several rounds of validation and refinement before they actually match operational reality as demonstrated in Fig.~\ref{fig:method}.

The iterative process shown in the right half of Fig.~\ref{fig:method} \mbox{requires} that the network operator prototype all major dialect deployment and management steps and experimentally evaluate how these steps may impact the performance assurance results derived from the security and cost analyses. In particular, potential impacts on the performance of authorized traffic flows and on the effectiveness of network monitoring tools should be measured and analyzed to guide selection and refinement of dialect implementation and management. Some negative impacts can be overarching and not correctable by implementation or management changes alone; such issues necessitate re-designing some of the dialect's basic elements, i.e., returning to Step~1 of the methodology. An example of an implementation limitation, which requires a step back to design, would be timing requirements of devices that are no longer satisfied due to the addition of deep packet inspection and security checks. In this case, the design required fast functioning security algorithms and packet inspection to not add excessive time delay.

Inspiration for the the dialect design process begins with the selection of a security concern by an organization, leading to the selection of a security mechanism and its associated options. This design process is similar to SP 800-37 Rev 2. Risk Management Framework for Information Systems and Organizations select stage~\cite{ross18}.  In this stage, one or more derivatives are designed and created for testing. Examples of desirable security objectives include confidentiality and/or authenticity. For this research, we selected authenticity as the primary security objective.

\subsection{Downgrade Attacks Against TLS}
Downgrade attacks are relatively well-known and are documented in Common Vulnerabilities and Exposures (CVEs)\footnote{CVE-2016-0800, CVE-2018-12404, CVE-2018-19608, CVE-2018-16868, CVE-2018-16889, and CVE-2018-16870.}~\cite{ncc19,aviram16,downgrade}. If such an attack was to occur between an SDN switch and controller, control channel messages would be vulnerable. For example, an attacker could send flow table updates to the switch in order to alter routing rules, or alter data statistics sent from a switch to the controller. 
As SDN devices become geographically spread, the use of TLS will be more important compared to physically secured data centers that may not use TLS, and consequently the implications of downgrade attacks on TLS for SDNs gain prominence. There are a variety of downgrade attacks available~\cite{downgrade}, and even the mitigations introduced with TLS 1.3 do not prevent a version downgrade (such as to TLS 1.2)~\cite{ncc19}.

\begin{figure}[!htbp]
    \centering
    \includegraphics[width=90mm]{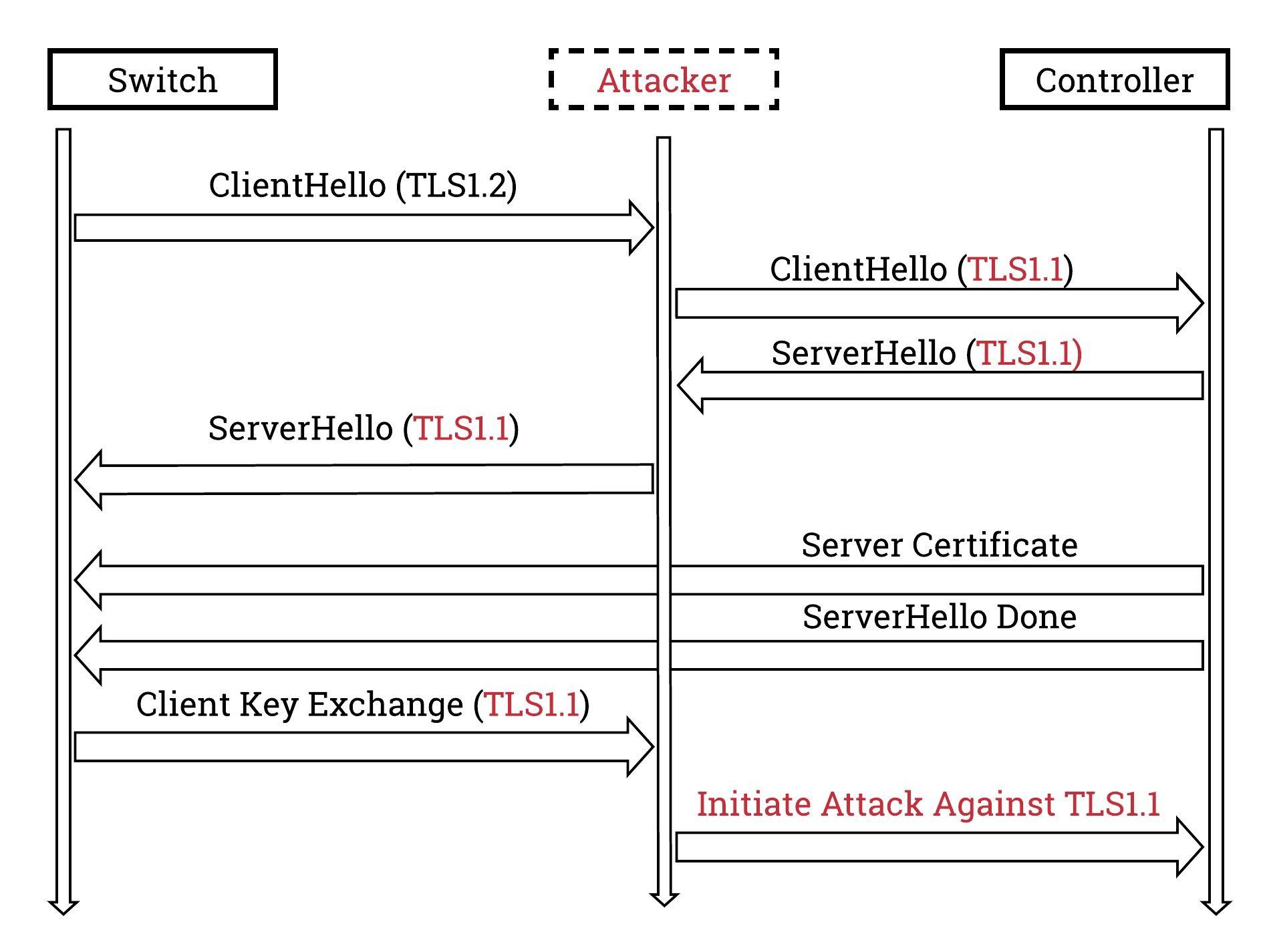}
    \caption{Illustration of TLS downgrade attack. Actions performed by the MitM attacker are indicated in red. These messages are unauthenticated in TLS1.2.}
    \label{fig:cipher-downgrade}
\end{figure}

As an example, consider a version downgrade on TLS 1.2. Since the initial ClientHello handshake message is  unauthenticated, a man-in-the-middle (MitM) attacker can downgrade the TLS version. This is illustrated in Fig. \ref{fig:cipher-downgrade}.  The attacker could simultaneously downgrade the advertised ciphersuite options, leading to the  establishment of a TLS session that is susceptible to a range of attacks. 

\subsection{Related Work}
\label{sec:related}

We summarize related work on three fronts: (i) efforts that explore a similar approach of system \mbox{specialization}, (ii) efforts that aim to improve SDN security using alternative approaches, and (iii) efforts for mitigating TLS downgrade attacks. 

{\bf System Specialization.} Broadly speaking protocol dialecting and code debloating~\cite{trimmer18,quach18,koo19}, can be classified as system specialization techniques. Notably, two recent papers, one on application program interface (API) specialization (Shredder~\cite{mishra18}) and the other on side-channel authentication~\cite{perazzone18}, illustrate the utility of system specialization in providing an additional layer of security protection and bolstering a defense-in-depth model.  

The Shredder system reduces the attack surface for an attacker to perform code reuse attacks by limiting the available function calls of an API (e.g., the binary of a C++ library) to those actually used by applications of an organization~\cite{mishra18}. The input parameter vector and value of each parameter for permitted function calls can be further restricted to those combinations used by the same applications. That and similar code debloating efforts~\cite{trimmer18,quach18,koo19} effectively restrict code execution paths and values of variables, based on results from static analysis or dynamic profiling of the code with respect to usage by target applications. In comparison, our work is focused on protocol-level semantics, including authenticity of control messages and selection of protocol parameters, and thus does not require code analysis or profiling. This is possible by the use of wrappers or randomized packet fields of the OpenFlow protocol.

Electromagnetic side-channel authentication~\cite{perazzone18} is a similar concept to protocol dialecting in that it also aims to add a new layer of authentication to the original communication protocol without modifying data packet contents. This security method relies upon physical layer modifications to add identity features at the source and verification at the receiver of the communication. Particularly worth noting is its ability to add a unique end-point fingerprint that is operating on a \emph{totally separate} plane from that of the original protocol. In comparison, the design explored in this paper aims to modify protocol behaviors within the same plane.

{\bf SDN Security.} Much prior research on SDN security has been focused on verifying the consistency and accuracy of new flow rules at the controller before deploying them to switches (e.g.,~\cite{veriflow,reitblatt12}). Our work is complementary to this line of research by providing control channel protection so that (i) rules that have been properly verified will be less likely to be modified without detection in transit before they reach their destination switches, and (ii) network state updates received by the controller are more likely to be authentic and thus, the rule verification performed at the controller is more likely based on accurate network state information.

Interestingly, the ConGuard work~\cite{xu17} shows that an adversary can exploit race conditions among different software threads of a deployed SDN controller to disrupt network operation and even crash a controller. They demonstrated this simply by generating a specific sequence of network events from a compromised host without needing to breach the SDN devices or  control channel. 
Some of those attacks leverage an absence of authentication on the control channel, while others operate via replay attacks. Our protocol dialecting solution enables early authentication while also preventing replay attacks.

{\bf Mitigation of Downgrade Attacks.} A countermeasure for TLS downgrade attacks was proposed in RFC~7507~\cite{rfc7507}. The solution requires the client to signal a special ciphersuite value called ``Fallback Signaling Cipher Suite'', indicating to the server that the server should generate an alert and abort the connection if its highest supported protocol version is higher than the version indicated by the client. Unfortunately, this solution cannot defend against the (MitM) attacker considered in this paper. The attacker can remove the special signaling value from the initial client message.  

Additional guidelines for TLS software implementations aiming to reduce potential vulnerabilities to protocol downgrade attacks are provided in RFC 7525~\cite{rfc7525}, but while such provisions reduce vulnerability, they have been shown not to eliminate it~\cite{ncc19}. This research offers a layered method for verifying that no changes have been made to TLS packages sent between SDN control channel devices, especially during the initial handshake; this includes authenticating relayed ciphersuite settings or version number, through a defense-in-depth approach.

\section{Dialecting~ OpenFlow}\label{sec:method}


In this section we describe in detail how we make use of the six-step methodology presented in Section~\ref{sec:criteria} to design, implement, and evaluate two different OpenFlow dialecting solutions. We refer to these solutions as \emph{OpenFlow derivatives} in the remainder of the paper. Each solution provides a different method for including additional layers of security. The methods we select include (i) a wrapper modification and (ii) a direct packet modification within the constraints of the protocol.

In addition to the two high-level criteria introduced in the beginning of Section~\ref{sec:criteria}, we also seek to minimize changes to the normal communication process of the SDN control channel devices in order to minimize additional network management complexity required for deploying the proposed OpenFlow derivatives. To this end, we consider a class of derivatives, which we term \emph{inline} solutions, that re-purposes existing OpenFlow messages for new security checks. These checks are then performed by the inline proxies with the potential for switch or controller code modification as well.

Furthermore, we make two simplifying design assumptions as follows: (i) clocks of SDN  devices are always synchronized within 100 milliseconds, and (ii) pre-shared keys are securely deployed \emph{a priori} between the controller and each switch. The clock synchronization can be achieved by deploying the Network Time Protocol (NTP) in the network or attaching GPS hardware to each SDN device. The requirement of pre-shared keys clearly introduces additional network management complexity. However, we observe that it occurs only once for each control channel device pair. Moreover, while pre-shared keys are assumed for simplicity in testing purposes, a suitable key exchange and management could be instituted via another dialecting derivative.

\subsection{Protocol Analysis}

Protocol analysis consists of a thorough examination of the standard operation of the base communication protocol (i.e. OpenFlow) for the purpose of creating protocol derivatives. We follow an information security approach in this analysis, identifying baseline adversarial capabilities and protocol goals. 

{\bf Threat Model}. We consider an attacker who can monitor all communications, delay, intercept, replay, and replace arbitrary control messages between the controller and switches. We assume a strong threat model to simulate a network spread out across geographical areas and to not limit potential advantages of the attacker. We assume that the attacker's purpose is not to instigate denial-of-service attacks, but to breach the SDN control channel. We further assume that the attacker has no means to directly compromise software, storage, and data structures running on the controller and switches without first compromising the control channel. This also means that we assume no SDN device has been compromised directly on the network.

While this threat model under-estimates the scope and severity of some existing and future attacks against the SDN control channel, it supports a proof-of-concept for protocol dialecting as a promising new direction for protecting SDN networks, while not being over-aiming for an all-encompassing solution.  Thus, this work serves as an initial step toward a comprehensive solution based on protocol dialecting. 

\begin{figure}[htbp]
\begin{subfigure}{.48\textwidth}
  \centering
    \includegraphics[width=90mm]{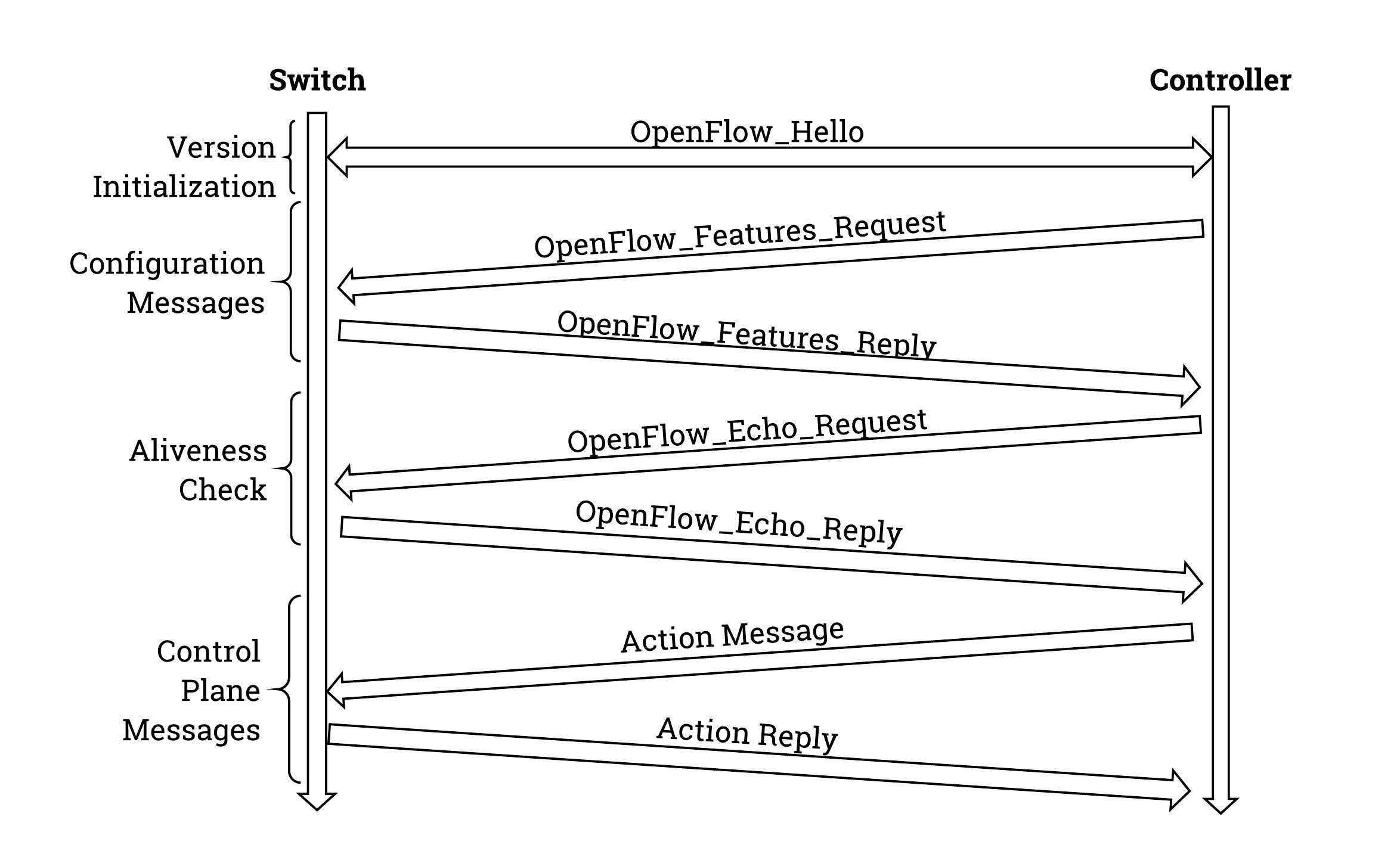}
    \caption{OpenFlow without TLS. 
    }
    \label{fig:without_TLS}
\end{subfigure}%

\begin{subfigure}{.48\textwidth}
  \centering
    \includegraphics[width=90mm]{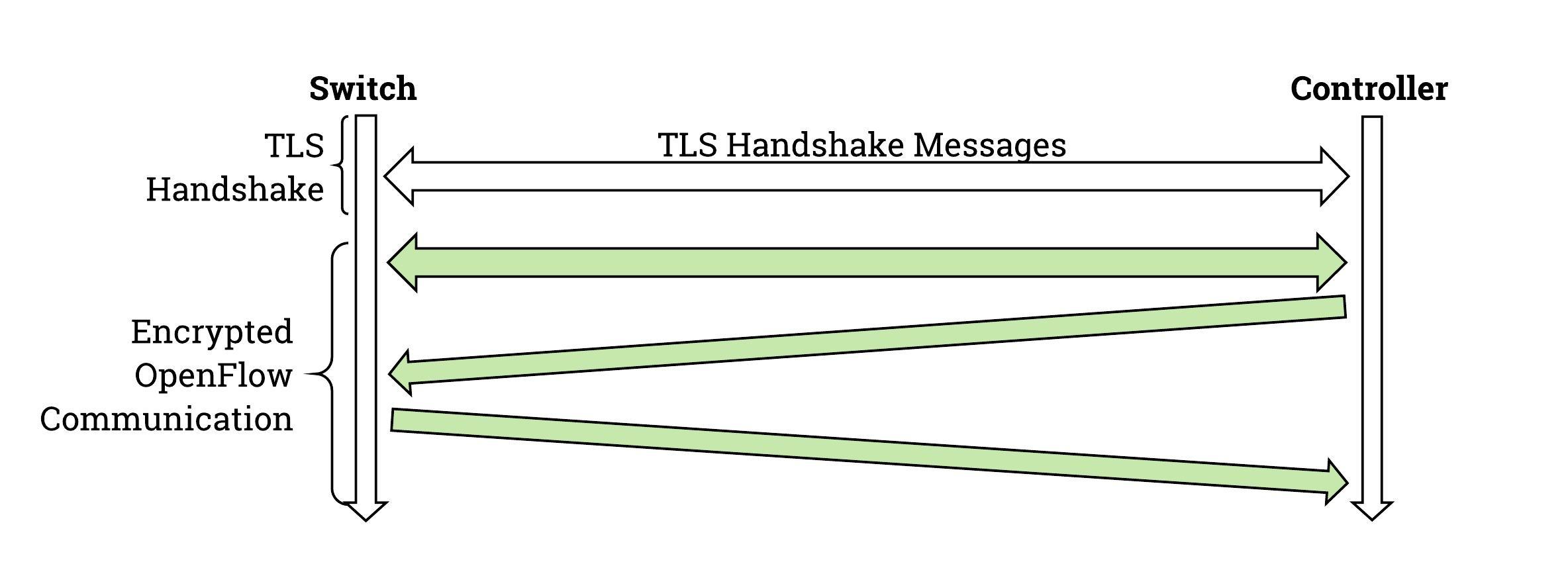}
    \caption{OpenFlow with TLS.}
    \label{fig:with_TLS}
\end{subfigure}
\caption{OpenFlow Communication comparison with/without TLS. Encrypted communication messages follow structure of messages without TLS.}
\label{fig:openflow}
\end{figure}

{\bf OpenFlow Message Features and Protocol Patterns}. 
In the next step, while developing in-line solutions, we focus on identifying (i) the prevailing protocol patterns between the controller and a switch and the critical messages used in each pattern, and (ii) header features of these messages with potentially reusable fields. 
We pay particular attention to the  latter; for example, a field that is normally random can allow for modification without requiring changes to the protocol specification. Such fields may be modified, still within the bounds of the OpenFlow standard, to provide packet space for inclusion of authentication tags. 

OpenFlow protocol communication is conducted between the controller and each participating switch,  in \emph{an identical point to point fashion}. The primary two communication patterns of each such connection are illustrated in Fig.~\ref{fig:openflow}. When TLS is not being used, the protocol begins with (unauthenticated) OpenFlow Hello messages sent non-interactively from the switch and controller~\cite{of1.5}. These messages are used to determine the highest version of OpenFlow that each party can support.  Next, the controller will send a Features\_Request message to establish configuration parameters for the switch, such as buffer size and number of tables supported~\cite{of1.5}.  Following channel configuration, the switch and controller will conduct a liveliness check based on a pre-defined interval (e.g. 5 seconds)using a ping-like mechanism via a pair of Echo\_Request and Echo\_Reply messages.

When TLS is enabled, the switch will begin communications with the controller by performing a TLS handshake.  Following the handshake, the switch and controller will perform the same setup procedures conducted as above, but this time the messages are encrypted, as illustrated with shaded arrows. 

We make two observations regarding the OpenFlow messaging patterns. First, a switch-specific exchange of Hello messages is required for the controller to establish communication with each switch, and another switch-specific periodic exchange of Echo messages is used for maintaining the connection. Given their importance and pervasiveness, these messages are good candidates for dialecting to provide additional security features. 
Second, when TLS is used, content of OpenFlow messages are meaningful at the receiving end  only after the outer TLS encapsulation is properly removed. This implies that if dialecting of OpenFlow messages is carried out before TLS functionality at the sending end, i.e., based on plain-text fields of OpenFlow messages, TLS and/or SDN parts of switch software must be dialected. Dialecting kernel level switch software may incur significant management complexity in a network deploying switches of different models and worse, from different vendors. This motivates a wrapper variant of dialecting to operate for every message sent between devices, which does not depend on the OpenFlow message format.

\begin{figure*}
{\centering
  \includegraphics[width=5.0in]{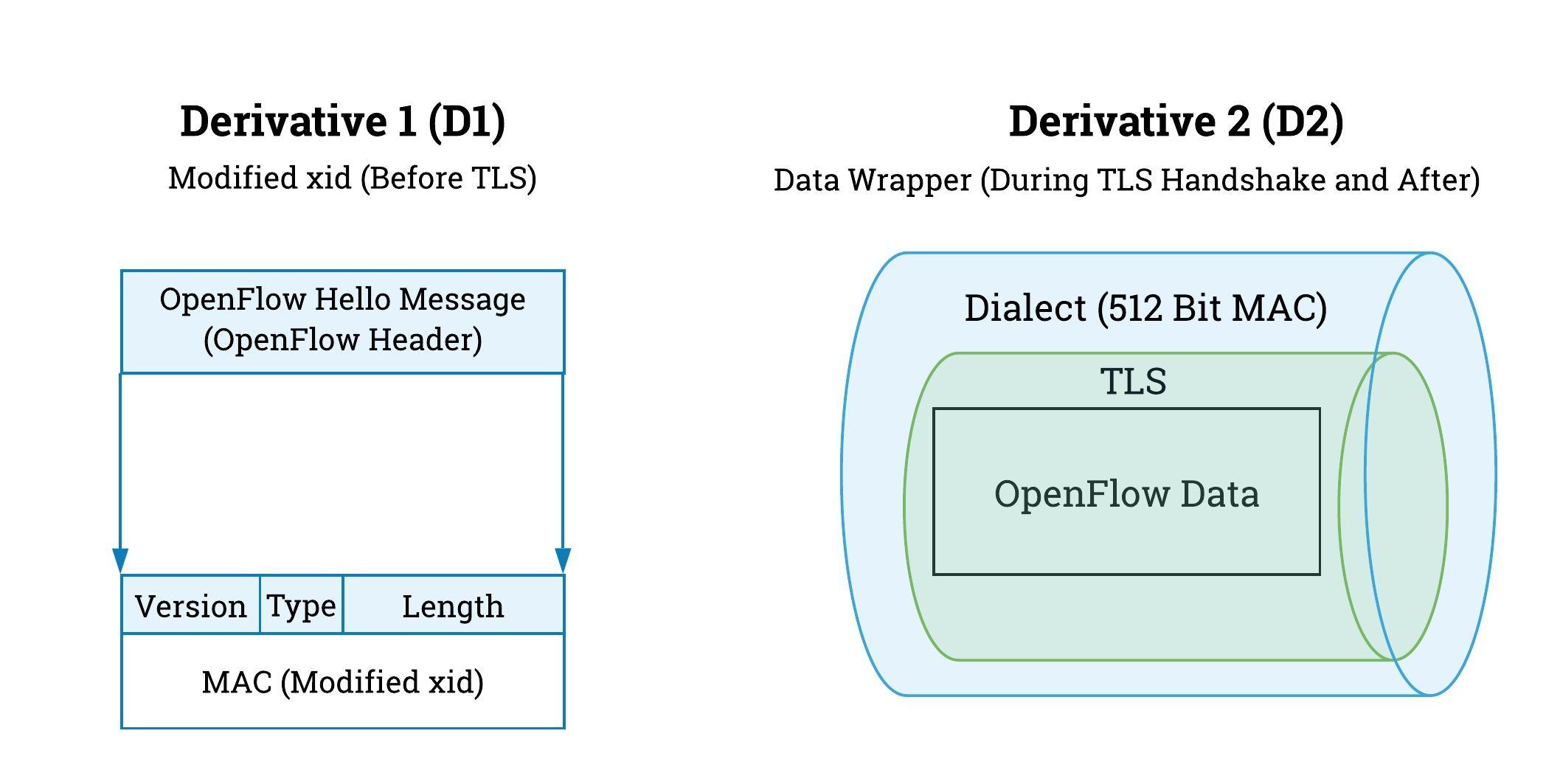}
  \caption{Illustration of Dialecting of Existing OpenFlow Messages to Support the New Derivatives. D2 outer (blue) shading indicates the derivative while the inner (green) shading indicates the TLS wrapper. Adapted from~\cite{sjoholmsierchio}.} 

  \label{fig:derivative_overview}
}
\end{figure*}

\subsection{Dialect Design}


The dialect design stage affects all future stages.  For this reason, it is important to consider how the choices made will impact key stages such as security and implementation, and revisit design choices during those stages. Derivative design choices important to this stage include security objectives,  cryptographic algorithms, and key generation and management. 

From the protocol analysis stage, we identify  two possible derivatives to add message-level authentication that is independent of TLS security features. These are implemented sequentially in our testing. 

{\bf Derivative 1 Design}. Derivative 1 (D1) modifies data within the bounds of the protocol, but re-purposes the 32-bit Transaction ID (xid) field of the initial OpenFlow Hello exchange as a hash-based message authentication code (HMAC). This allows for immediate authentication of packets  before TLS is established.

The D1 Hello message format is shown in the left part of Fig.~\ref{fig:derivative_overview}.  The transaction ID field of each OpenFlow Hello message is normally randomly generated at both the controller and switch, thus amiable to re-purposing.  The HMAC verification is performed by the proxies at both the controller and switch to prevent malicious commands and data from reaching the SDN devices. 

The available 32-bit transaction ID field is well below recommended MAC length levels~\cite{nist800107}. This issue is unsurprising when considering dialecting with re-purposed existing header fields; however, it raises a design question of ensuring security even under a limited bit field that can easily be brute-forced. To accomplish this, we effectively limit the HMAC lifespan by "killing off" keys from memory after a pre-determined interval so that the MAC verification cannot take place after the security window is exceeded.
More specifically, we utilize a Hash-based Key Derivation Function (HKDF)~\cite{rfc5869} to derive two uni-directional keys from the switch-controller pair-specific pre-shared key and the BLAKE2b\cite{2013blake2} hash algorithm for computing HMACs.
New keys are then derived sequentially in a ratcheted fashion (the current key is used to derive the next key), with each key having a lifespan of 1 second. Our calculations to justify selection of a 1 second key lifespan are based on average collision resistance of
$2^{32 \text{ bits} / 2}$ and an average Python test time to create a 32-bit MAC of $10^{-4}$ seconds. This yields an average of $6.5$ seconds necessary to find a collision.

It should be noted that D1 cannot be used indefinitely without potentially interfering with the OpenFlow standard.  When the controller or switch exchange multiple messages with the corresponding \texttt{more} flag set, follow-up messages are required to utilize the same xid field~\cite{of1.5}.  Under D1, this field is overwritten for each message and, since the contents of the message are different, the xid field will contain a different MAC for the subsequent messages.  This is not an issue during setup of the initial connection, however, as each message does not use the \texttt{more} flag.

{\bf Derivative 2 Design}. While D1 is native to the OpenFlow protocol, it only adds protection to the initial communication between the controller and switch for the reasons stated above. Conceivably, we could expand the protection while using a similar strategy to dialect Echo messages. However, with Derivative 2 (D2), we chose to develop a more comprehensive solution covering \emph{all} OpenFlow messages. There is no limitation on using both D1 and D2 to reduce initial communication forgeries and per-packet protection following the initial OpenFlow Hello.

Furthermore, D2 is designed to allow inter-operation with TLS. To avoid management complexity associated with dialecting kernel level switch software, we define D2 as message wrapper as follows. A 512-bit HMAC tag is appended to encrypted data packet before sending, as illustrated in the right part of Fig.~\ref{fig:derivative_overview}. The communicating ends accept an OpenFlow message as valid only if it passes the HMAC check.  D2 is protocol agnostic and can be added to a variety of protocols (e.g. various versions of TLS or when TLS is not enabled) because it does not manipulate any internal fields or data in the packets.

We use the HKDF and BLAKE2b algorithms for key derivation and MAC computation for both D1 and D2, with two differences. First, in D2, the HMAC length is increased to 512 bits. The length can be further increased based on network security requirements. Second, under the extended MAC tag length we no longer bound the key lifetime, but still require per-message keys as well as unique keys for each sender. We use a unidirectional sequence number (i.e. one sequence number per sender, incremented per message) as part of the hash input to combat replay attacks. 

For instance, once the TLS handshake has completed, the switch and controller initialize independent (derivative) sequence numbers.  These numbers are included in the HKDF function and incremented after each message.  In the event an attacker attempts to replay a previous message, the receiving side will attempt verification using the correct sequence number and detect an error, resulting in a dropped packet. In our experiment we set the receiver to drop the packet and connection completely if an error associated with a validation code was detected. 
Since re-establishing the switch-controller connection requires relatively little time,
we select to recreate the setup of the switch versus simply dropping such error packets and keeping the connection. 

 A summary of the options selected for D1 and D2 are shown in Table~\ref{table:security_selections}. 

\begin{table}[htbp]
\caption{Derivative D1 \& D2 Security Selections}
\begin{center}
\begin{tabular}{|l|l|l|}
\hline
\textit{\textbf{Derivative Security Options}} & \multicolumn{2}{l|}{\textit{\textbf{Selected Options}}} \\ \hline
Information Security Objective & \multicolumn{2}{l|}{HKDF} \\ \hline
\multicolumn{1}{|l|}{\multirow{2}{*}{Replay Protection}} & D1 & 1-Second MAC Lifetime \\ \cline{2-3} 
\multicolumn{1}{|c|}{} & D2 & Unidirectional Sequence Number \\ \hline
Algorithm & \multicolumn{2}{l|}{HMAC} \\ \hline
Sub-Algorithm & \multicolumn{2}{l|}{BLAKE2b} \\ \hline
\end{tabular}%
\label{tab1}
\label{table:security_selections}
\end{center}

\end{table}

\subsection{Security and Cost Analysis}
The security and cost analysis stage is an intermediate analysis step to ensure that these objectives have been met. An organization's requirements determine if the added security is sufficient, as well as the impact on overhead and system delay. The addition of security solutions causes the use of additional processing power, memory, and time. If a given organization's security requirements and cost constraints are not met during this intermediate analysis, then the dialect designer must return to stage 1. If the security and cost objectives have been met, then the designer may continue in the process to determine the implementation method. 

We defer the analysis of our design choices to its own section (Section~\ref{sec:analysis}).

\subsection{Implementation Method}

We observe that there are important considerations in selecting an implementation method for our OpenFlow derivatives. One option we considered was to borrow from code debloating work~\cite{trimmer18,quach18,koo19} and extend some of the existing automated tools that manipulate binaries directly to support adding new functionality to existing OpenFlow software. This approach looks attractive at first because it would enable rapid deployment of derivatives without requiring much code development effort by the operator. However, upon closer examination, we found major limitations with it. The tools operate at granularity of loops and function calls, thus lacking the ability to rapidly prototype, enforce high-level security policies, and be compatible with a variety of protocols.

{\bf Deployment via Policy Enforcement Proxies}. Based on these observations, and to enforce high-level security policies such as select  versions of TLS or a certain ciphersuite, and to speed up evaluation of the derivatives, we choose to create a new software component called the policy enforcement proxy (PEP). This method is compatible with Mininet as our independent system for testing and measurement without using multiple physical systems, and allows for testing to occur only with one machine.

\begin{figure}[htbp]
    \centering
    \includegraphics[width=90mm]{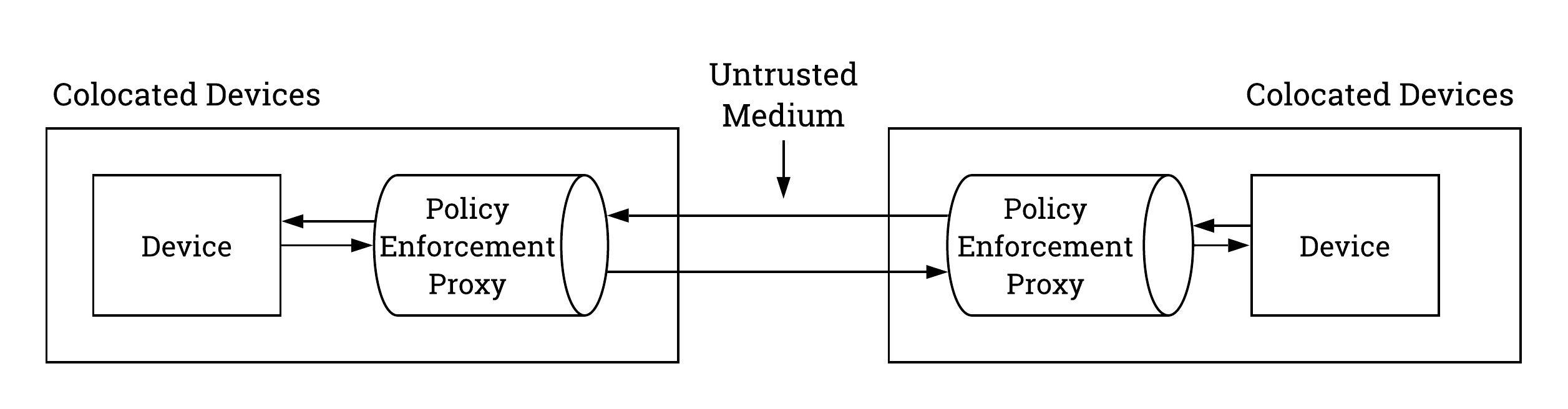}
    \caption{Proxy Implementation Method.}
    \label{fig:proxymethod}
\end{figure}
\begin{figure}[ht]
    \centering
    \includegraphics[width=90mm]{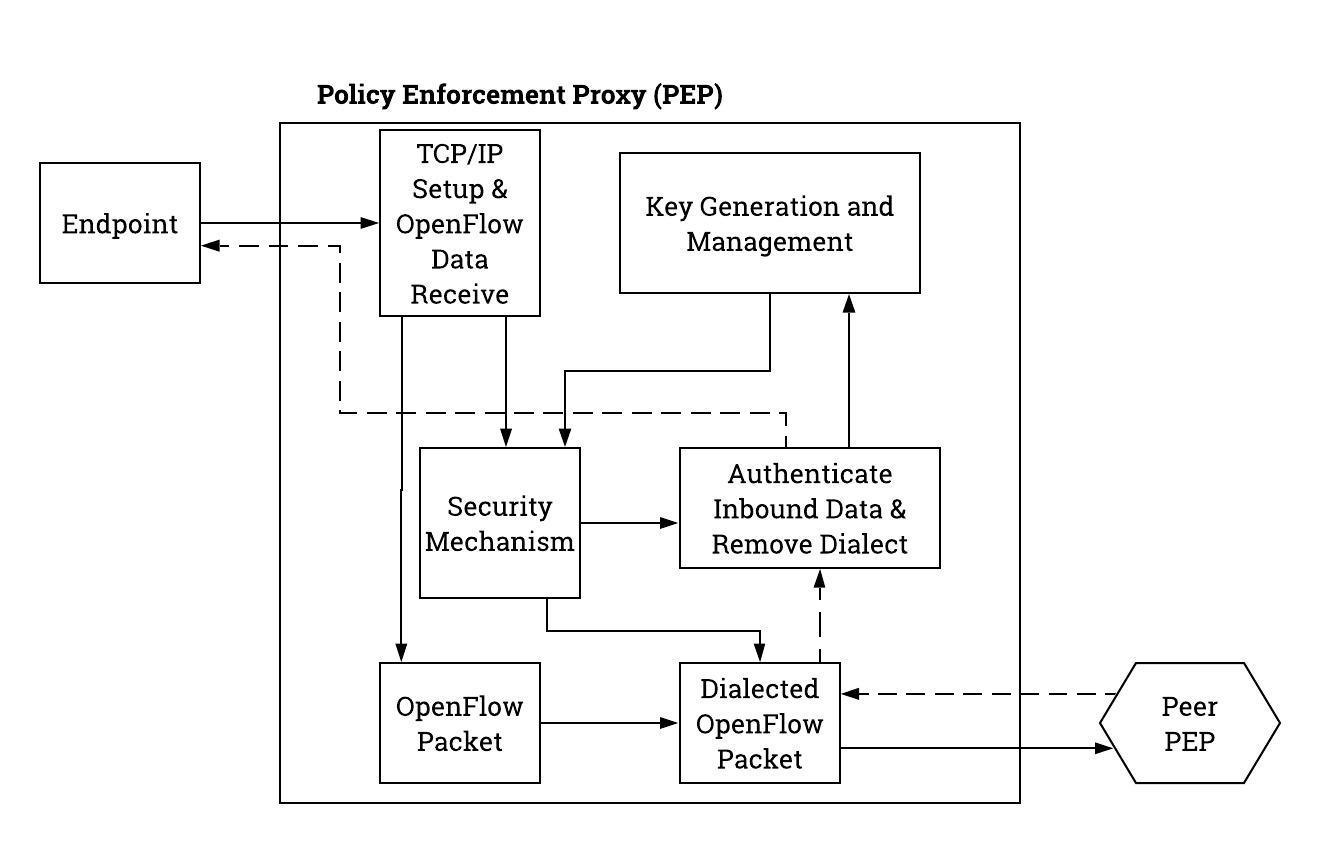}
    \caption{Policy Enforcement Proxy Design. This figure demonstrates the logical setup of the proxy and how data is managed between functions.}
    \label{fig:Security_enforcement_design}
\end{figure}

The PEP implementation method is shown in Fig.~\ref{fig:proxymethod}, with more detailed D1 and D2 functionality expanded in Fig.~\ref{fig:Security_enforcement_design}. A pair of proxies sit between each switch and the controller, and consequently do not have access to the TLS keys used by the switch or controller.  For each OpenFlow message, the proxy  assigned to the sending device intercepts and dialects messages, without decrypting them, as shown in Fig.~\ref{fig:derivative_overview}. It  then sends the dialected message to the peer proxy. The receiving proxy verifies the derivative and performs other checks as required by the security policy.  For this work, we installed at the PEPs two sample security policy rules, regarding the TLS protocol version and ciphersuites, respectively. We see other possibilities for security policy rules, e.g., rate limiting of flow requests and link state updates from a switch as a countermeasure to denial-of-service attacks. We also expect that the PEPs can be extended to enforce and regulate other protocols and features besides OpenFlow and TLS, as discussed in Section~\ref{sec:discussion}. 

\subsection{Dialect Management}
This stage involves the support systems and methods that will allow for PEPs to operate long-term and in day-to-day operations. Attributes important to dialect management include ease of management, installation difficulty, and scalability.

To maximize security, design of OpenFlow derivatives should require an independent key management system. In this work, we assume that TLS runs between the switch and controller, while D2 is proxy-to-proxy. This separation means that compromise of the end device does not necessarily imply loss of derivative keys (on the proxy), or vice-versa. Conceivably, to reduce the deployment cost, the operator may choose to reuse the same physical infrastructure as for storing TLS keys. In such a scenario, we recommend to minimally use virtualization techniques to establish a logical partition between the different key computing, storage, and communication resources used for the derivative and TLS. Even with such precautions, it should be noted that security of reuse of such a physical infrastructure for both the derivative and TLS is dependent on the absence of compromise of the end devices. 

\subsection{Implementation Testing}
Implementation is the last stage of our protocol dialect process. Implementation testing consists of testing security properties and the availability of the system or protocol with the included dialect. Before this testing can occur, control tests of normal operation are performed and later used as a testing baseline following dialect integration. The communication latency and other types of overhead added due to the addition of a dialect are then measured and calculated to ensure that availability of the system meets the organization's requirements. For clarity of presentation, we defer the details of our testing efforts to Section~\ref{sec:evaluation}.
\section{Security Basis}
\label{sec:analysis}

In this section we take a closer look at the security of D1 and D2, based on the design choices discussed in Section~\ref{sec:method}. 
We did not conduct a formal security proof of the composition of D1 and D2, leaving that to future work, but provide justification arguments for the design security through comparison to prior work.

\subsection{Formalizing Design Choices}
\begin{figure}[htbp!]
    \centering
    \includegraphics[width=60mm]{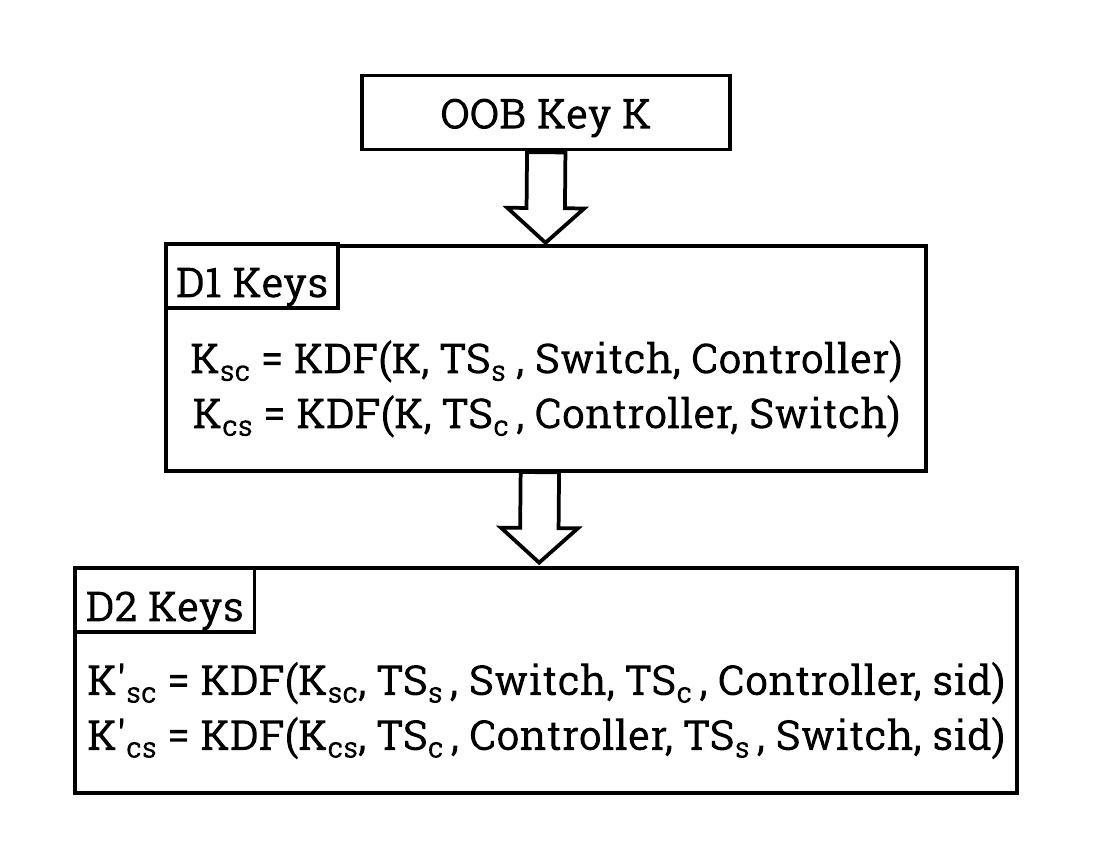}
    \caption{D1, D2 Key Derivation. $\mathtt{KDF}$ represents the key derivation function, $\mathtt{TS_a}$ indicates the timestamp for party $\mathtt{a}$, $\mathtt{K_{ab}}$ indicates the symmetric key used uni-directionally between $\mathtt{a}$ and $\mathtt{b}$, and $\mathtt{sid}$ represents the session identifier which we calculate as \mbox{$\mathtt{sid= D1\_MAC\_Switch||D1\_MAC\_Controller}$}.}
    \label{fig:keyderivation}
\end{figure}

\begin{figure}[htbp]
    \centering
    \includegraphics[width=90mm]{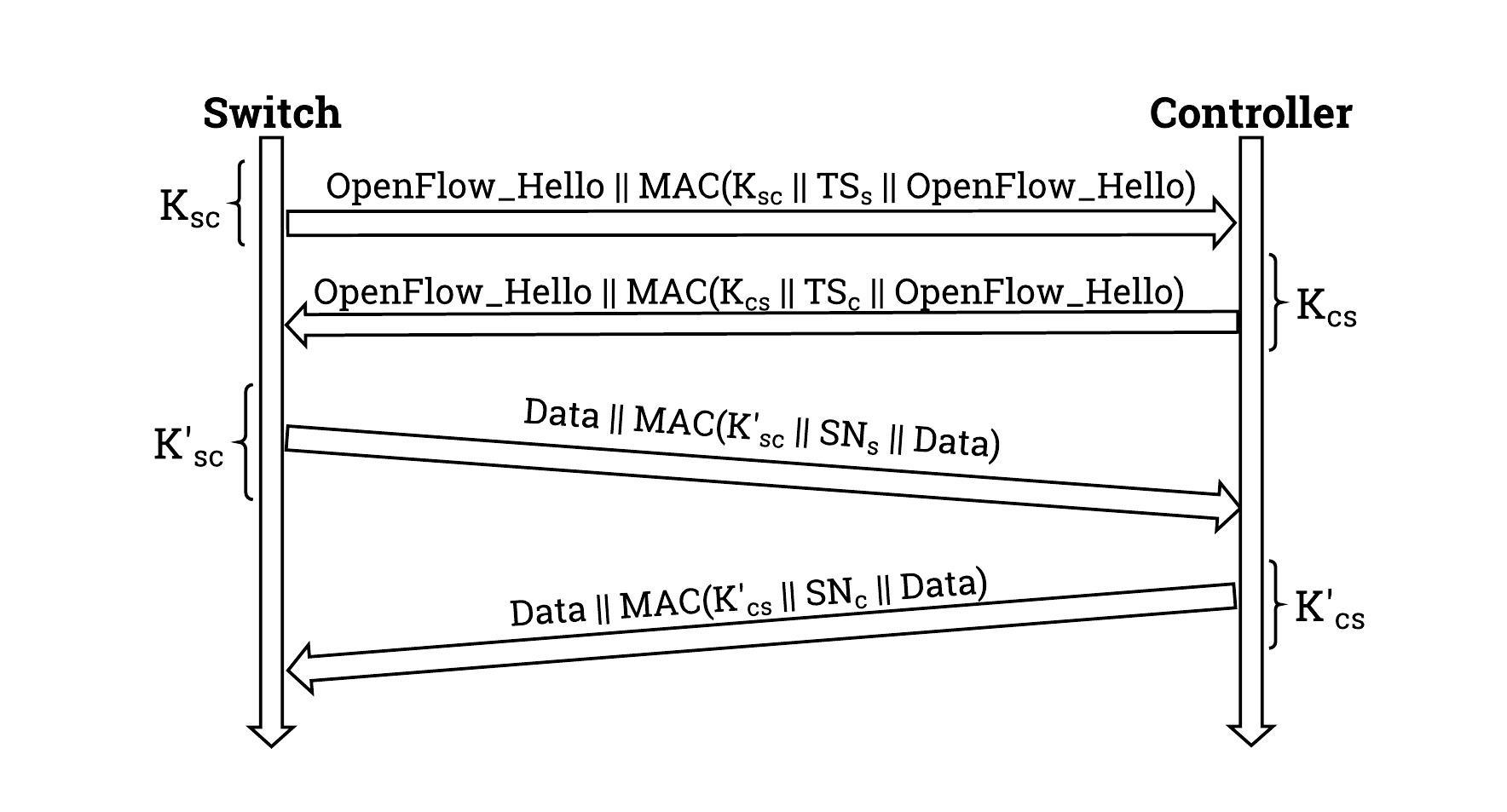}
    \caption{D1, D2 Key Usage. Horizontal flows indicate (potentially) non-interactive messages such that ordering is not enforced. Angled flows indicate interactive messages with enforced ordering. $\mathtt{SN_a}$ indicates sequence number for party $\mathtt{a}$, $\mathtt{TS_a}$ indicates the timestamp for party $\mathtt{a}$.}
    \label{fig:keyusage}
\end{figure}

In the following discussion, we assume use of both D1 and D2 in a sequential format. Prior to implementation of D1 and D2, a pre-shared symmetric key, generated at random, is established and delivered out-of-band (OOB).\footnote{This may occur as an OOB key distribution. However, there is the potential to embed a full key exchange in a prior run of D2; we leave this as future work.} It is assumed that keys used between a controller and switch are not reused for any other controller/switch pair. Following the initial key distribution, each side generates uni-directional keys (UDK) for D1 and D2 according to Fig.~\ref{fig:keyderivation} and Fig.~\ref{fig:keyusage}, where $K_{ab}$ indicates the symmetric key used to protect data from $a$ to $b$ and $ || $ indicates concatenation.

As seen in Fig.~\ref{fig:keyusage}, the proxies will initially exchange D1 messages potentially non-interactively (the OpenFlow protocol does not mandate an order to the Hello messages)~\cite{of1.5}.
The freshness value incorporated in our design is based on implicit timestamps (rounded to seconds). This has implied assumptions that the controller and switch have the same clock time, which is reasonable with embedded hardware clocks. 
Although such synchronization constitutes a potential point of failure, the timestamp could be computed based on the completion of the TCP handshake to limit clock de-synchronization.

Packets arriving during the same second should pass D1 MAC verification, while those arriving outside of the time window will fail verification.
Note that we do not employ timestamp values simply to inhibit replay attacks; as noted earlier, the MAC length is limited in D1 by the 32-bit header field and as a result is liable to forgery attacks. Consequently, it is necessary to limit the viable lifespan of the MAC, which is performed via the key lifespan window.   
Any packet received outside of this window will have a different key used in the D1 MAC calculation step.  Therefore, if an attacker attempts to replay any messages, they must be replayed within the same time window in order to pass the MAC verification step.

Note that we use the transcript D1 MAC values of the initial flows to provide continuity to the communication between D1 and D2, essentially employing these as a session identifier.  This links protection of the initial D1 security layer to later layers.

\subsection{D1 Security}

The security goal of D1 is to establish that the proxies involved possess the correct keys and perform MAC verification of the OpenFlow messages. Thus D1 acts, in part, as a mutual authentication protocol with associated data. We reference \cite{rogaway2009authentication} for in-depth analysis methods which may be applied to such composed protocols with associated data. 
An attacker's goal of breaking entity authentication during D1 is largely dependent on breaking the security of the MAC, necessitating the use of a short MAC lifespan through validity windows, due to the MAC output length used during D1.

\begin{figure}[htbp]
\begin{subfigure}{.48\textwidth}
  \centering
    \includegraphics[width=88mm, trim=30 0 0 0]{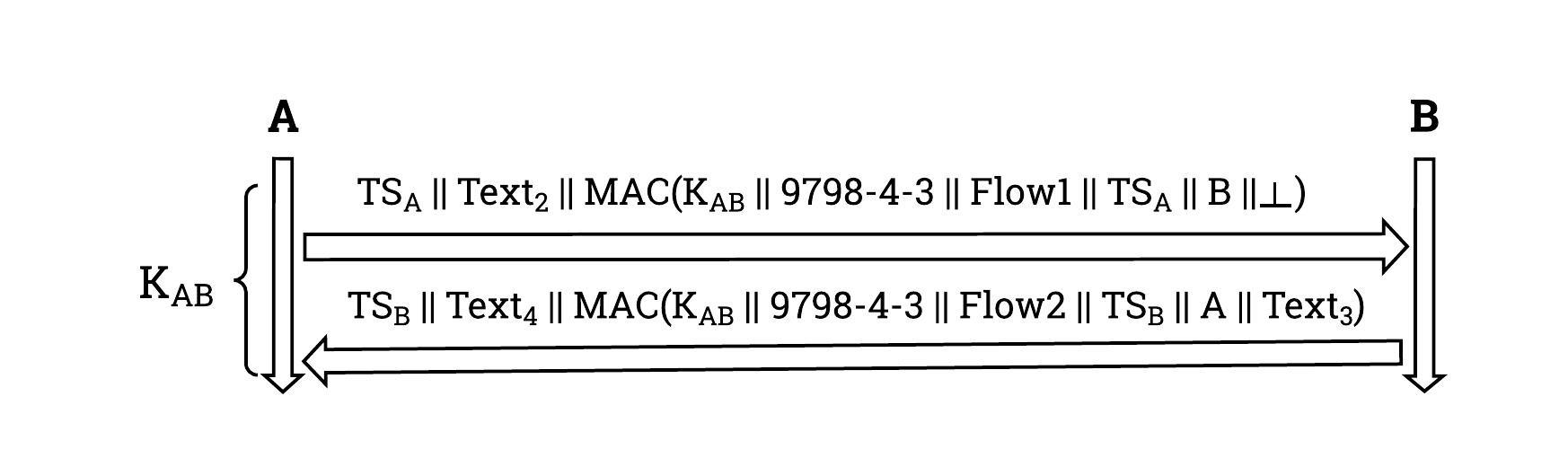}
    \caption{Repaired ISO/IEC 9798-4-5.2.1.}
    \label{fig:repaired_flow}
\end{subfigure}%

\begin{subfigure}{.48\textwidth}
  \centering
    \includegraphics[width=88mm, trim=0 0 25 0]{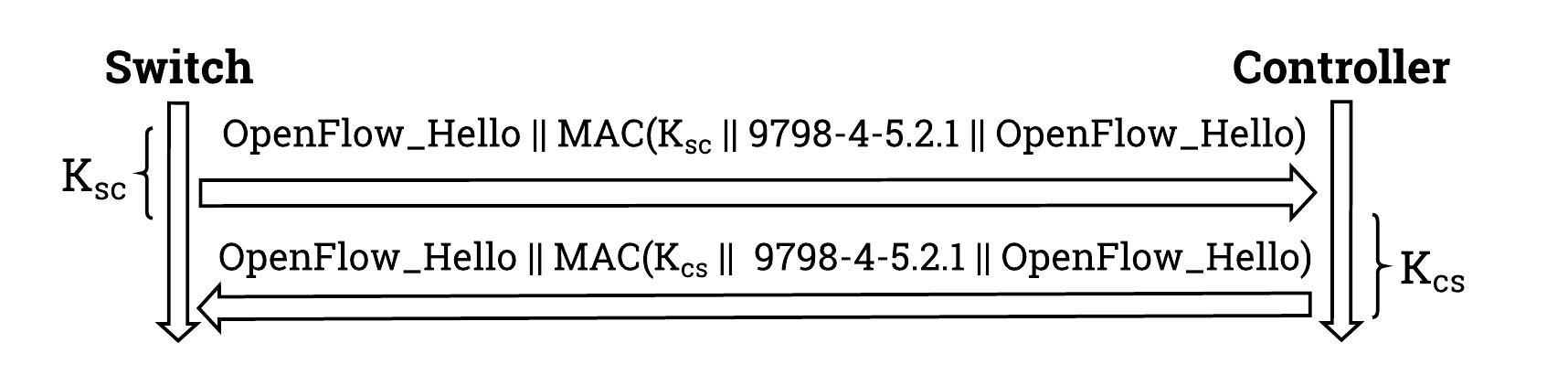}
    \caption{D1.}
    \label{fig:D1_flow}
\end{subfigure}
\caption{Comparison of repaired ISO/IEC 9798-4 Mechanism 5.2.1 two-pass mutual authentication protocol to D1. Note that ISO/IEC 9798-4:1999 Mechanism 5.2.1 uses a generic check function, with a MAC as one possible instantiation; it is represented here as a MAC.   The $\mathtt{Flow1}$ (resp. $\mathtt{Flow2}$) value corresponds to a protocol flow position indicator introduced by \cite{basin2013provably}. 
    $\mathtt{TS_S}$ (resp. $\mathtt{TS_C}$) is
    incorporated into the derivation of
    $\mathtt{K_{sc}}$ (resp. $\mathtt{K_{cs}}$).
    $\mathtt{K_{sc}}$ and $\mathtt{K_{cs}}$ also address the directional separation.
    $\mathtt{OpenFlow\_Hello}$ messages constitute the
    authenticated text $\mathtt{Text_x}$ (note that in
    ISO/IEC 9798-4:1999 Mechanism 5.2.1, it is not required that all
    text fields are the same). }
    \label{fig:9798_compare}
\end{figure}

As a justification of the authentication properties of D1, we draw to comparison with a previously analyzed protocol, the repaired ISO/IEC 9798-4:1999 Mechanism 5.2.1 two-pass mutual authentication protocol (referred to as ISO 9798-4-3 by Cremers et al.) \cite{basin2013provably,ISO9798}. The comparison is shown in Fig.~\ref{fig:9798_compare}. We specifically note and justify the following. 

\paragraph{Check Function} ISO/IEC 9798-4:1999 Mechanism 5.2.1 specifies the use of a cryptographic check function, with reference to those given in ISO/IEC 9797. This includes our choice of a MAC algorithm. From this point on we will assume that the cryptographic check function is a MAC.     

\paragraph{Uni-Directional Keys (UDK)}\label{par:first}
Cremers et al. \cite{basin2013provably} incorporate identities into the check function of ISO/IEC 9798-4:1999 Mechanism 5.2.1 to show direction of flows, hence avoiding reflection attacks arising from use of a single symmetric key. D1 embeds this in the derivation process of the UDKs.  Each key derivation contains identifiers for the sender and the receiver.  This prevents messages sent by the switch to be accepted by the switch, e.g. a reflection attack.  Furthermore, the control plane of an SDN limits communication from a switch only to the controller (switches will not communicate with other switches). 
When UDKs are used, such as described above, the identities are no longer a necessary input to the cryptographic check function~\cite{ISO9798}.

\paragraph{Text Fields} $Text_x$ values in Fig.~\ref{fig:9798_compare} correspond to the OpenFlow Hello messages sent by D1. Unlike in the original ISO/IEC 9798-4:1999 Mechanism 5.2.1, we do not allow for variable selection of data for these fields: it is pre-determined. Consequently, we also avoid the associated potential protocol flow syntax errors \cite{basin2013provably}.  

\paragraph{Timestamps}
The repaired ISO/IEC 9798-4:1999 Mechanism 5.2.1 explicitly sends timestamps ($TS_x$) which it uses as inputs to the MAC. D1 uses a form of implicit timestamps, which are used in key derivation for the MAC vs.  input into the protocol data fields. By using implicit timestamps, D1 conforms as closely as possible to the OpenFlow standard. Moreover, implicit use is enabled as the switch and controller have shared knowledge of the timestamps.   

\paragraph{Flow Ordering}
The repaired ISO/IEC 9798-4:1999 Mechanism 5.2.1 protocol~\cite{basin2013provably} uses a flow tagging value to ensure messages are correctly interpreted according to their order in the protocol, e.g. ``$\mathtt{Flow1}$'' and ``$\mathtt{Flow2}$''. We require that D1 proxies enforce ordering, where the protocol flow generated by the switch will always be interpreted as the first flow, while the protocol flow from the controller will be held to be the second.

\paragraph{Protocol Version Confirmation}\label{par:last} As in \cite{basin2013provably}, we explicitly send the authentication protocol version (``$\mathtt{9798-4-5.2.1}$'') used in D1. 

With consideration of the above points a)--f), we map D1 to the repaired ISO/IEC 9798-4:1999 Mechanism 5.2.1, which achieves mutual authentication of party $A$ and $B$ \cite{basin2013provably}. Thus, D1 acts as a pre-shared key mutual entity authentication protocol for the subsequent D2 derivative in addition to providing authentication of the additional data fields, namely the OpenFlow Hello messages.

\subsection{D2 Security}

The PEP logic requires that D1 is completed before starting D2, and therefore mutual authentication has already taken place.  On the completion of D1, the switch and controller proxies create UDKs used for D2.
The security goal of D2 is to provide an authentication layer for the remainder of the controller/switch session. In this case, the authenticated ``data'' is in fact the actual OpenFlow packet. Keys for this derivative are derived as presented earlier and shown in Fig.~\ref{fig:keyderivation}. 
In particular, D2 keys are derived from those of D1 and are computed over the transcript of D1 messages sent, thereby binding D2 to the earlier entity authentication and ensuring both parties commit to the transcript of D1.

D2 makes use of a 512-bit MAC that allows for a longer MAC lifetime than D1. Additionally, D2 incorporates sequence numbers into the MAC calculation step as seen in Fig.~\ref{fig:keyusage}, which increment for each new packet sent under a given key. Sequence numbers are thus uni-directional to each sender. Notably, this limits the potential for replay of messages by an attacker, which is of particular concern in an SDN due to the controller sending configuration commands to the switch.  If an attacker was able to replay an old configuration command, then the switch could be induced to start routing traffic improperly or the update may result in erratic network behavior.  

\section{Experimental Evaluation}
\label{sec:evaluation}

\begin{figure}[ht]
    \centering
    \includegraphics[width=87mm]{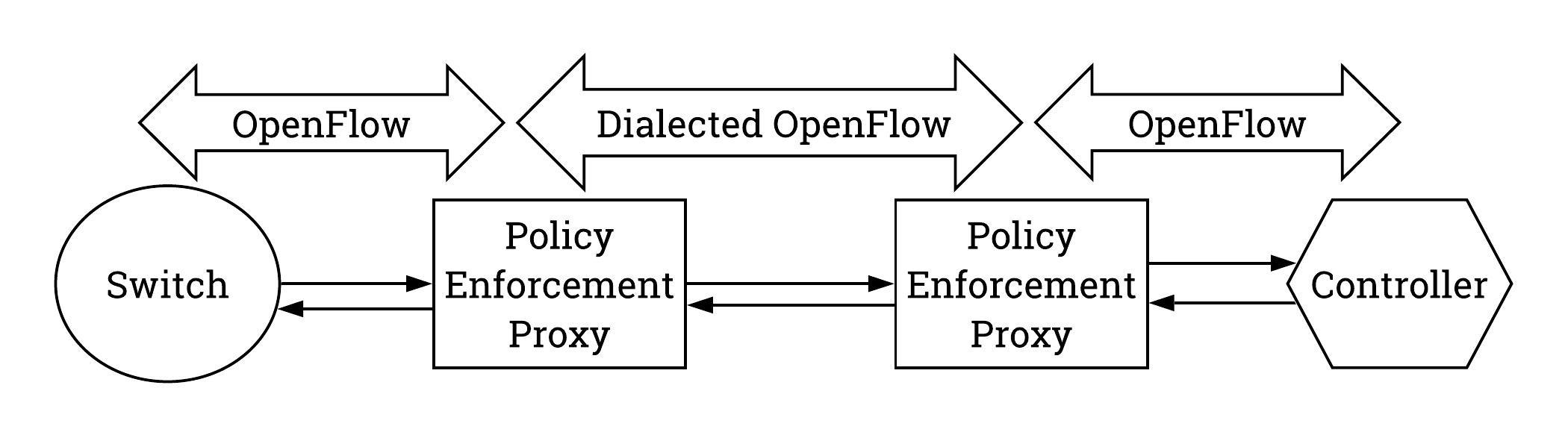}
    \caption{Mininet Testbed Typology}
    \label{fig:Mininet_Experiment_Topology}
\end{figure}

In this section we present results from our experimentation on the D1 and D2 derivatives in a testbed built with the Mininet SDN emulation software~\cite{mininet}. 
The testbed topology with the PEPs is shown in Fig.~\ref{fig:Mininet_Experiment_Topology}. The raw packet data as well as timing statistics were captured using the Wireshark tool~\cite{wireshark}.

\subsection{Mitigation of Downgrade Attacks}

First, we perform a simulation of an attack scenario, validation that the operation of the derivative  against a TLS version downgrade attack. 
We assume that an attacker can modify packets at will between the controller and the switch. Therefore, it is the responsibility of the receiving proxy to enforce that the modified packet by the attacker does not make it to the end point device. 

\begin{figure}[ht]
    \centering
    \includegraphics[width=87mm]{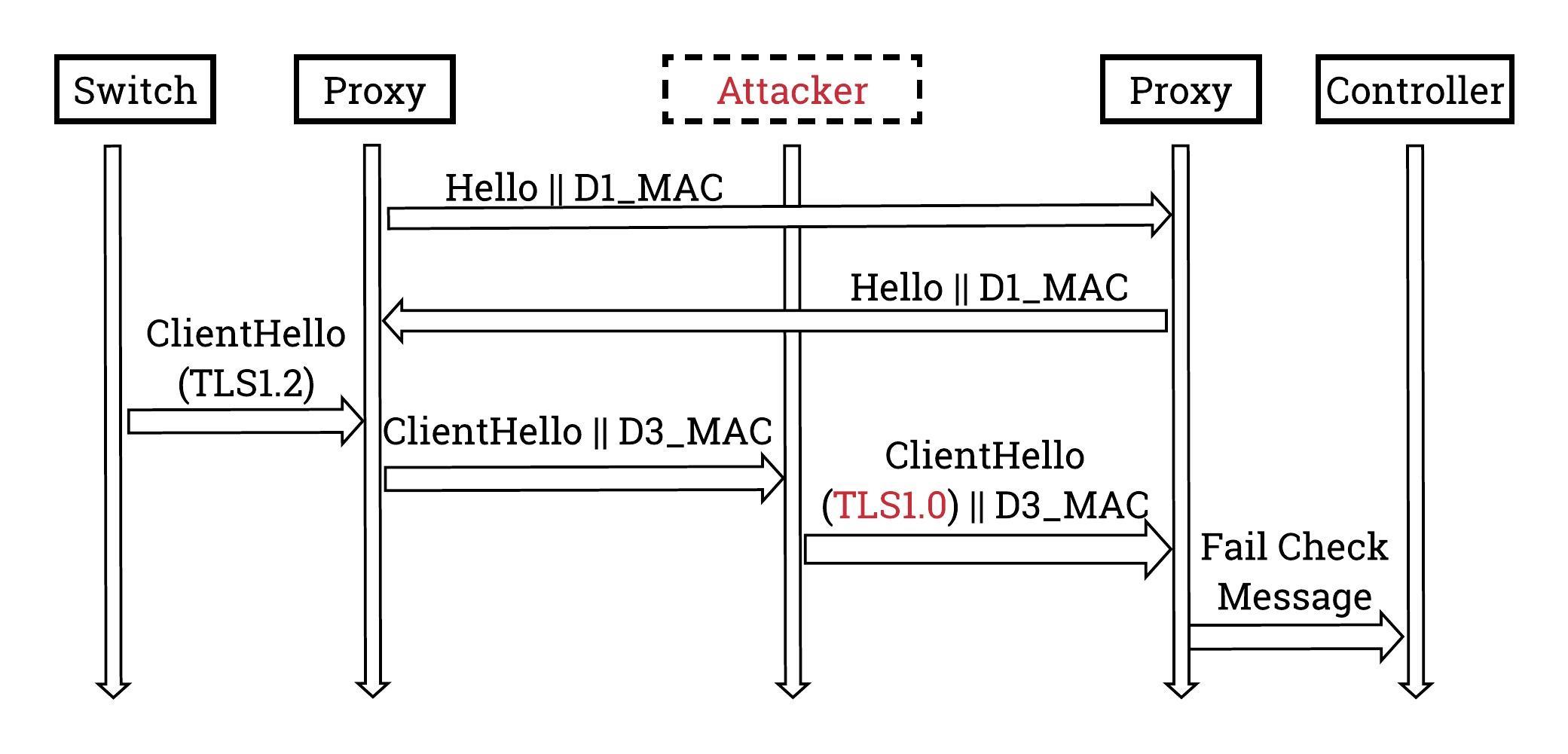}
    \caption{Timing Diagram of Downgrade Attack Simulation. The initial Hello messages exchanged by the proxies are OpenFlow Hello messages.  D1\_MAC / D2\_MAC corresponds to the MAC performed during that derivative as seen in Fig.~\ref{fig:keyusage}.}
    \label{fig:downgrade_dialect_exp}
\end{figure}

The method for the attack simulation is shown in  Fig.~\ref{fig:downgrade_dialect_exp}. In this experiment we modify the data in the first message of the TLS handshake (i.e., the switch's ClientHello) from TLS version 1.2 to 1.0 before the message is delivered to the PEP for the controller. Two versions of the attack were carried out. The first version consisted of modifying the TLS version number without changing to the MAC tag.  The second version modified the TLS version number while also changing the MAC tag (i.e. to simulate the attacker "guessing" the MAC).  In both cases, the receiving PEP  rejected the attack packet by flagging the wrong D2 MAC value.
The PEP then sent an alert to the controller so that the controller could terminate the TLS handshake.

We note that it is straightforward to add logic to the PEPs to ensure that the ciphersuite advertised by each party conforms to the IETF guidelines~\cite{rfc7507,rfc7525}. For brevity, we omit such experiments.  


\subsection{Estimation of Communication Overhead}

After verifying that the derivatives can be used to stop a TLS downgrade attack, we conduct a series of experiments to quantify the communication overhead from using the D1 and D2 derivatives.  
Baseline experiments without utilizing the PEPs were performed, and to examine the impact of TLS on communication latency, some experiments were also conducted without turning on TLS. 
Specifically, we conducted a total of 90 tests for three different types of experiments as summarized in Table~\ref{table:experiment_runs}.

\begin{table}[htbp]
\caption{Summary of Overhead Measurement Tests}
\begin{center}
\begin{tabular}{|l|l|}
\hline
\textit{\textbf{Experiment Type}} & \textit{\textbf{Number of Runs}} \\ \hline
Baseline & 30 \\ \hline
TLS Enabled & 30 \\ \hline
TLS Enabled and Using D1 \& D2 & 30 \\ \hline
\end{tabular}
\label{table:experiment_runs}
\end{center}
\end{table}

By default, TLS is disabled in Mininet which allows for the observation of traffic and the protocol. Therefore, we performed timing first without the PEPs added to the system. The average time to establish a connection between a controller and switch was measured to be 5 seconds. To perform control testing with TLS required a generic setup and the establishment of certificates~\cite{ggee_2014}. Enabling TLS with Wireshark also requires the use of the preinstalled ovs-testcontroller as well as an OVSSwitch. The average time to establish a connection with TLS was measured to be 9.33 seconds.

In the next step of the experiment, 
delays with and without TLS were measured from the first OpenFlow Hello to the first OpenFlow Echo Request message. This measurement was taken because it demonstrates the complete process to setup a new connection between the switch and controller. The OpenFlow Echo Request message signifies that all setup actions have completed and the SDN devices are switching to a liveliness check state~\cite{of1.5}.

A summary of the average connection time measurements and the average percentages of latency increase due to TLS and the derivatives are shown in Table~\ref{table:exp_results}. In both dialecting cases, PEPs where used. The overhead added by the derivatives without using TLS was minimal. Using TLS alone significantly increased the connection latency, by an average of about 87\%.  Then, deploying the derivatives on top of TLS further increased the latency by  an average of approximately 22\%. This larger overhead increase (compared to the the case without using TLS) can be attributed to the larger packet sizes of TLS and corresponding MAC processing. Still, we consider the connection latency increase over TLS to be moderate, unlikely to impact network availability.

\begin{table}[htbp]
\caption{Summary of Average Connection Latency}
\resizebox{87mm}{!}{%
\begin{tabular}{|l|l|l|l|l|}
\hline
\textbf{TLS } & \textbf{Using} & \textbf{Average} & \textbf{Overhead} \\ 
\textbf{Enabled} & \textbf{D1 \& D2} & \textbf{Latency} & \textbf{} \\ 
\hline
No & No & 5.00 sec. & Baseline \\ 
\hline
No & Yes & 5.01 sec. & 
0.0\% \\ 
\hline 
Yes & No & 9.33 sec. & New Baseline \\
&&& (87\% beyond no security) \\ 
\hline
Yes & Yes & 11.37 sec. & 22\% beyond TLS \\
&&& (127\% beyond no security)\\ 
\hline
\end{tabular}%
}
\label{table:exp_results}
\end{table}

\subsection{Reproducing the Experiments}

The virtual machine, scripts, packages, and Mininet used to perform all experiments can be found open-source at https://tinyurl.com/teqedac. The virtual machine provided contains bash scripts to perform all the experiments described in this paper.
The administrator account password is set to "default" for modification and running all experiments. Running each experiment only requires commands in a command line interface. 

\section{Discussion}
\label{sec:discussion}

In this section we discuss some of the limitations and potential extensions of this work.

\subsection{Limitations}

While we have focused on downgrade attacks, it is unknown whether or not this solution is effective in mitigating other types of attacks against TLS in SDNs. As discussed in Section~\ref{sec:related}, we observe that dialecting proxies can infer the associated network events from received OpenFlow messages even in an encrypted form~\cite{baset07}. The intelligence gained from such inference could be helpful in mitigating other forms of attacks targeting TLS.

Dialect deployment cost evaluation in this work constitutes initial testing only, with attention to qualitative arguments.
To properly evaluate the cost and compare it to that of the standard operation of OpenFlow and TLS necessitates a rigorous quantification of major cost factors. This includes the management complexity associated with generating and protecting the pre-shared keys and  re-configuring incumbent network monitoring tools. We leave this quantification to future work.  

Finally, our experimentation uses a simplistic testbed topology. While this topology allowed for quantification of relative latency overhead, there may be factors that only manifest in a more sophisticated topology. For example, when the controller needs to talk to many switches at the same time, the proxy on the controller could become a performance bottleneck. Consequently, performance evaluation under large topologies should be considered before deployment. Use of a low level programming language such as C/C++ to prototype the performance critical parts of the proxy logic, is another future extension.

One could ask about the security-cost trade-off of the D1 and D2 derivatives. After all, if TLS requires a key exchange for each new connection between a given switch and controller, are we not doubling the set-up efficiency cost to also obtain a shared derivative key and cannot we not simply upgrade to a newer version of TLS? 
In answer to this, we highlight the point-to-point and defense-in-depth approach to our solution. 
\paragraph{Initial Set-Up Cost} Once an initial pre-shared key is obtained, a new D1 key is not required for subsequent sessions, but rather can be derived from the last D2 key in use. Thus the derivative keys are consistently ratcheted forward over time. The first D1 key in use could potentially be derived for a TLS pre-shared key. In this case, the first controller-switch communication would be protected only by TLS, while in subsequent sessions D1 would protect the TLS handshake. Alternative methods for obtaining a pre-shared D1 key are also possible, allowing for a great deal of flexibility. 
\paragraph{Reliance on TLS} While TLS relies on a public key backbone infrastructure, our derivatives do not, and are adaptable to different infrastructures, such as identity-based cryptography, or underlying protocols. Depending on organizational need, TLS may not even be the ideal underlying protocol. Moreover, updating all SDN devices to, say, TLS 1.3, may not be feasible dependent on network agility. Once a protocol is implemented, especially over a wide, distributed network, small modifications to data packets (such as in D1) may be significantly more viable than a syndicated upgrade.
\paragraph{Defense-in-Depth}
Protocol dialecting is explicitly about adding a security layer to an existing protocol, versus a new stand-alone solution. In this case, we look at an added layer of authenticity to protect initial handshakes and mitigate downgrade attacks. Network goals vary, and it is not always feasible to  update to a specific protocol, even if there is one that provides for all organizational needs. Consider a case in a larger network with various network proxies: in a traditional design, security options are limited to end-to-end security (no packet visibility at proxies) or proxy-to-proxy security (devices must trust proxies not to alter data). With a dialecting approach, one could add end-to-end authenticity, while accepting proxy-to-proxy TLS to allow for visibility. 
Thus, our efficiency analysis is of note in that it demonstrates the relative feasibility of such layered solutions.

\subsection{Extensions}

\subsubsection{A General Security Framework Based on PEPs}

We observe that protocol dialecting is particularly suited for identifying attackers in a zero trust model network that assumes all devices, even hosts inside the enterprise firewall, may be malicious~\cite{kinder2010}. That is, all traffic on the network is assumed to be unsafe unless properly verified. With a protocol dialect, the operator can deem only the dialect traffic to be trusted between corresponding devices. In this light, a PEP can be viewed as a potential platform for performing all security functions such as access control, intrusion detection, and content filtering, at different levels of granularity (per protocol, per device, or per service). A dialect presents a method to add security to protocols that have limited security options and features, or are lacking particular security guarantees desired by administrators.

Using PEPs to safeguard enterprise security policy can be more flexible than the current device centric solutions because it allows the operator to separate security policy definition and enforcement from other detailed device configuration tasks and furthermore, rapidly deploying new security policies. In other words, PEPs can be the basis of a unified policy framework for enterprise security.  

We recognize that many technical challenges remain to turn this vision of policy framework into effective solutions. Among them are (i) the need for a flexible dialecting solution that can accommodate various protocols and enterprise applications in one integrated system, and (ii) the need for an operator-friendly policy definition language and associated automation tools for deploying and configuring PEPs. 
Furthermore, we provide an early caution against overuse of PEPs, such as in careless combinations of various  security policies and dialecting. While PEPs can enable the benefits mentioned above, overuse can lead to a single point of failure as well as presenting a vulnerability to the system through misconfiguration or contradictory policies. Endpoint security -- on devices such as the SDN controller and switch  -- remains critical. Notably, dialecting does not rely on PEPs, but can be facilitated by them.

\subsubsection{New Derivative for Key Management}

Our design of the two dialecting derivatives (D1 and D2) assumes that each switch side PEP possesses a pre-shared secret with the controller side PEP.  To allow dynamic replacement of the shared secret, we envision an option of a third derivative (D3) that introduces a new type of OpenFlow Experimenter message~\cite{sdn-msg} to carry out key management functions between two communicating PEPs.

D3 may operate inline with current OpenFlow sessions. Since it is desirable to separate PEP operation from that of TLS, i.e. TLS and dialect keys are independent and stored on different devices (SDN device vs. PEP), D3 key negotiation cannot take place directly between the switch and controller. 
In one solution, 
the controller side PEP and switch side PEP may use the respective controller and switch as relays for the negotiation.
Alternatively,
D3 may operate independently from OpenFlow sessions, establishing a channel between each pair of communicating PEPs that may be used to negotiate new keying material.

Note that it D3 could potentially rely on the same certificates used in TLS for initial negotiation, and thereafter ratcheting keys forward in a continuous key agreement (similar to how D2 keys are derived from D1 keys). While this does not supply key separation from TLS at first use, the security benefits grow over time; compromise of the long-term private key later on would effect TLS session handshake negotiations, but not the derivative. We leave such key exchange derivative options and their analysis as future work.

\section{Conclusion}
\label{sec:conclusion}

In this research, we have shown it is feasible to leverage protocol dialecting to add another layer of per-message authentication into the OpenFlow protocol that is  independent of TLS.

A key benefit of this additional message authentication is that through its enforcement during the TLS handshake, the SDN control channel is more robust to TLS downgrade attacks. Additionally, the dialecting solution incurred a moderate increase of communication overhead, less than 22\% over TLS, as measured on a Mininet testbed. 

We view  protocol dialecting as a general and effective solution for deploying and enforcing enterprise security policy on a per-protocol basis. 
Protocol dialecting also supports the industry-led movement toward zero trust access control models~\cite{kinder2010}. This work raises interesting new research questions such as how to develop a cross-protocol dialecting framework and how to ensure that the resulting dialects continue to work with current network monitoring and analysis systems.


\bibliographystyle{IEEEtran}
\bibliography{protodialect}{}

\end{document}